\documentclass[journal,10pt]{IEEEtran}
\usepackage{amsmath}
\ifCLASSINFOpdf
  \usepackage[pdftex]{graphicx}
  \graphicspath{{../pdf/}{../jpeg/}}
  \DeclareGraphicsExtensions{.pdf,.jpeg,.png}
\else
  \usepackage[dvips]{graphicx}
  \DeclareGraphicsExtensions{.eps}
\fi

\usepackage{subfig}

\usepackage{amsmath,amssymb,amsthm,mathrsfs,dsfont}

\usepackage{cite}

\usepackage{tabularx}

\begin{document}

\title{A Contract-based Incentive Mechanism for Energy Harvesting-based Internet of Things}

%\author{
%\IEEEauthorblockN{
%Zhanwei Hou,
%He Chen,
%Yonghui Li
%~and Branka Vucetic%,~\IEEEmembership{Fellow,~IEEE}
%}\\
%\IEEEauthorblockA{School of Electrical and Information Engineering, University of Sydney, Sydney, NSW 2006, Australia\\
%E-mail:~\{zhanwei.hou,~he.chen,~yonghui.li,~branka.vucetic\}@sydney.edu.au}\\
%}

\author{
\IEEEauthorblockN{
Zhanwei Hou\IEEEauthorrefmark{1},
He Chen\IEEEauthorrefmark{1},
Yonghui Li\IEEEauthorrefmark{1},
Zhu Han\IEEEauthorrefmark{2},%~\IEEEmembership{Fellow,~IEEE},
~and Branka Vucetic\IEEEauthorrefmark{1}%,~\IEEEmembership{Fellow,~IEEE}
}\\
\IEEEauthorblockA{\IEEEauthorrefmark{1}School of Electrical and Information Engineering, University of Sydney, Sydney, NSW 2006, Australia\\
E-mail:~\{zhanwei.hou,~he.chen,~yonghui.li,~branka.vucetic\}@sydney.edu.au}\\
\IEEEauthorblockA{\IEEEauthorrefmark{2}Department of Electrical and Computer Engineering, University of Houston, Houston, TX 77004-4005 USA\\
E-mail:~\{zhan2@uh.edu\}}
}

\maketitle

\begin{abstract}
By enabling wireless devices to be charged wirelessly and remotely, radio frequency energy harvesting (RFEH) has become a promising technology to power the unattended Internet of Things (IoT) low-power devices. To enable this, in future IoT networks, besides the conventional data access points (DAPs) responsible for collecting data from IoT devices, energy access points (EAPs) should be deployed to transfer radio frequency (RF) energy to IoT devices to maintain their sustainable operations. In practice, the DAPs and EAPs may be operated by different operators and a DAP should provide certain incentives to motivate the surrounding EAPs to charge its associated IoT device(s) to assist its data collection. Motivated by this, in this paper we develop a contract theory-based incentive mechanism for the energy trading in RFEH assisted IoT systems. The necessary and sufficient condition for the feasibility of the formulated contract is analyzed. The optimal contract is derived to maximize the DAP's expected utility as well as the social welfare. Simulation results demonstrate the feasibility and effectiveness of the proposed incentive mechanism.
\end{abstract}

%\begin{IEEEkeywords}
%Internet of things, energy harvesting, energy trading, contract theory, incentive mechanism
%\end{IEEEkeywords}

\section{Introduction}
By connecting objects, physical devices, vehicles, animals and other items without human intervention, Internet of Things (IoT) has emerged as a new paradigm to enable ubiquitous and pervasive Internet connections \cite{Gubbi2013IoT}. Wireless sensing and monitoring service is one of the fundamental applications of IoT, which enables systems and users to continually monitor ambient environment.

One of the major hurdles for implementing the wireless sensing application is the limited lifetime of traditional battery-powered sensors, which are costly and hard to maintain. For example, frequent recharging or battery replacement is inconvenient in deserts or remote areas, and is even impossible for some scenarios, such as toxic environment or implanted medical applications \cite{Kama2015Wireless}. To tackle this problem, radio frequency energy harvesting (RFEH) has recently been proposed as an attractive technology to prolong the operational lifetime of sensors, enhance the deployment flexibility, and reduce the maintenance costs\cite{Kama2015Wireless,Niyato2016Novel}.

In this paper, we consider a RFEH-based IoT system consisting of a data access point (DAP) and several energy access points (EAPs). The DAP is in charge of collecting information from its associated sensors. The sensors are assumed to have no embedded energy supply, but they can harvest energy from radio frequency (RF) signals radiated by the surrounding EAPs to transmit the data to the DAP. In practice, the DAP and EAPs may be operated by different operators. To successfully motivate these third-party and self-interested EAPs to help charge the sensors, effective incentive mechanisms should be designed to improve the payoff of the DAP as well as those of EAPs.

Traditionally, the devices belonging to the same network with extra energy were assumed to voluntarily assist other devices, e.g., \cite{Medepally2010Voluntary}. However, this becomes no longer applicable for the considered system with self-interested third-party EAPs, as these EAPs tend to maximize their own benefits. In \cite{Chen2015stackelberg}, an incentive mechanism was designed for the system with the similar setup where monetary rewards were provided by the DAP to motivate third-party EAPs to assist the charging process. This process was referred to as ``energy trading''. The authors formulated the incentive problem as a Stackelberg game, where the DAP is the buyer for the RF energy and the EAP competes to sell energy to the DAP. Another auction-based incentive mechanism was developed and evaluated in \cite{ma2015distributed} for an alternative energy trading scenario with multiple DAPs and a single EAP. In these schemes, it was assumed that the EAP(s) will truthfully report some private information to the DAP(s), e.g., their energy costs and channel gains between EAPs and sensors. However, this assumption is not realistic. Since EAPs are selfish, in practice, an EAP may provide misleading information maliciously and pretend to be an EAP with better channel condition and/or higher energy cost to cheat for more rewards. A malicious EAP can succeed in cheating to get more benefits because of \emph{information asymmetry} in the energy trading process. Specifically, an EAP clearly knows its private information, such as its own energy cost and channel conditions towards sensors to be charged, which are generally hard to be known by the DAP. To address this issue, in this paper we will design an effective incentive mechanism to maximize the expected utilities of the DAP and EAPs by overcoming the information asymmetry. We are interested in addressing following questions without knowing the private information of EAPs: \emph{Which EAPs the DAP should hire, how much energy should be requested from the hired EAPs, and how many rewards should be given to the hired EAPs?}

To answer the above questions, we apply the well-established contract theory to design the incentive mechanism of the energy trading process in the considered RFEH-based IoT system. Contract theory is an powerful tool originated from economics to model the incentive mechanism under information asymmetry in a monopoly market. This problems is called ``adverse selection'' in contract theory \cite{Bolton2005Contract}. It has been employed to address incentive design problems in wireless communication areas, such as device-to-device (D2D) communications\cite{Zhang2015Contract} and cooperative spectrum sharing\cite{Duan2014Cooperative}. To the best knowledge of the authors, this is the first work that uses contract theory to design the incentives for the energy trading process in RFEH-based IoT systems.

In our design, the energy trading market is analogous as a monopoly market in economics. The DAP is the employer who offers a contract to each EAP. The contract is composed of a serious of contract items, which are combinations of energy-reward pairs. Each contract item is an agreement about how many rewards an EAP will get by contributing how much energy. Various heterogeneous EAPs are classified into different types according to their energy costs and instantaneous channel conditions. The EAPs are regarded as labors in the market, which will choose a contract item best meeting their interests. By properly designing the contract, an EAP's type will be revealed by its selection. Thus the DAP can capture each EAP's private information to a certain extent and thus overcome the issue of information asymmetry. During the design of the contract, we characterize the necessary and sufficient conditions for the contract feasibility, i.e, individual rationality (IR) conditions and incentive capability (IC) conditions. Subject to the IR and IC constraints, the optimal contract under information asymmetry is derived by maximizing the DAP's expected utility as well as the social welfare. Simulations validate the feasibility and effectiveness of the proposed incentive mechanism.

\section{System Model}
We consider one DAP and $N$ EAPs belonging to different operators, which are connected to constant power supplies. The DAP is responsible for collecting various data from several wireless-powered sensors within its serving region. Without embedded energy supplies, the wireless-powered sensors fully rely on the energy harvested from the RF signals emitted by the EAPs to transmit its information to the DAP. For simplicity, we consider that the RF energy transfer and information transmission are performed over orthogonal bandwidth. Since the EAPs are assumed to belong to different operators, they cannot collude with each other, i.e., energy trading among EAPs is not considered. For analytical tractability, time division-based transmission among sensors is adopted, i.e., there is only one active sensor during each transmission block. Hereafter, we refer to this active sensor as the information source. Besides, all the nodes in the system are assumed to be equipped with single antenna and operate in the half-duplex mode.

The DAP will offer a contract to effectively motivate the EAPs to charge its information source (i.e., the active sensor). In practice, the EAPs can be heterogeneous with different energy costs and instantaneous channel gains towards the information source. Obviously, there is asymmetric information between the DAP and EAPs. To be more precise, each EAP knows exactly its energy cost and channel status\footnote{Note that each EAP can estimate its channel towards the sensor via the uplink pilots sent by the sensor.}, which is, however, unknown to the DAP. To overcome this information asymmetry, the DAP will design a group of energy-reward contract items. Rewards can be monetary incentive or free offloading data between operators.

We consider that the energy-carrying signals sent by the EAPs are independent and identically distributed (i.i.d.) random variables with zero mean and unit variance. Note that no coordination between the EAPs is needed since independent signals are transmitted. All channels are assumed to experience independent slow and flat fading, where the channel gains remain constant during each transmission block and change independently from one block to another. The information source rectifies the RF signals received from the EAPs and uses the harvested energy to transmit its information. The time duration of every transmission block is normalized to one. So we use ``energy'' and ``power'' interchangeably hereafter. The amount of energy harvested by the information source during one transmission block can be expressed as
\begin{equation}\label{eq:Energy}
  E_{s} = \eta\sum\limits_{m=1}^{N} p_mG_{m,s},
\end{equation}
where $0<\eta<1$ is the energy harvesting efficiency, $p_m$ is the charging power of the $m$th EAP, and $G_{m,s}$ is the channel power gain between the $m$th EAP and the information source. Note that the noise is ignored in (\ref{eq:Energy}) since it is practically negligible at the energy receiver.

The harvest-use protocol is considered in this paper\cite{Krikidis2013harvest}. More specifically, the information source will use the harvested energy to perform instantaneous information transmission to the DAP. We consider a battery-free design which indicates that the sensor only has a storage device like supercapacitor to hold the harvested energy for a short period of time, e.g., among its scheduled transmission block. Hence the sensor exhausts all the harvested energy in each transmission block, so the sensor's energy storage device is emptied at the beginning of the transmission block. This battery-free design can reduce the complexity and costs of the sensors, which is particularly suitable for the considered IoT sensing applications and has been adopted by other applications \cite{Ju2014Throughput,Lu2015Wireless}. The transmit power of the information source is
\begin{equation}\label{eq:P_s}
  P_{s} = E_{s}.
\end{equation}
Then, the received signal-to-noise ratio (SNR) at the DAP is
\begin{equation}\label{eq:beta}
  \beta = \frac{p_s G_{a,s}}{N_0},
\end{equation}
where $N_0$ is the noise power at the DAP, and $G_{a,s}$ is the channel power gain from the information source to the DAP. Hence the achievable throughput (bps) from the information source to the DAP can be expressed by
\begin{equation}\label{eq:Ras}
\begin{aligned}
R_{sa} &= W \log_2 (1+\beta) \\
&=W \log_2 \left( 1+\frac{\eta G_{a,s}}{N_0} \sum\limits_{m=1}^{N} p_m G_{m,s} \right),
\end{aligned}
\end{equation}
where $W$ is the bandwidth. We define the received power contributed by the $m$th EAP as $q_m = p_m G_{m,s}$ and $\gamma = \eta G_{a,s}/N_0$ for notation simplicity. So (4) is simplified as
\begin{equation}\label{eq:Ras2}
R_{sa} = W \log_2 \left( 1 + \gamma \sum\limits_{m=1}^{N} q_m \right).
\end{equation}

In the following subsections, we will define the utilities of the DAP and EAPs as well as the social welfare.
\subsection{DAP's Utility}
Note that the aim of the DAP is to pay less rewards to the EAPs to achieve higher throughput. The DAP's utility can thus be defined as
\begin{equation}
U_{DAP} = W \log_2 \left( 1 + \gamma \sum\limits_{m=1}^{N} q_m \right) - c \sum\limits_{m=1}^{N}\pi_m,
\end{equation}
where $\pi_m$ is the money (or amount of free offloading data) paid by the DAP to the $m$th EAP for its corresponding contribution $q_m$, and $c$ is the unit cost of the DAP, which is normalized as $c=1$ without loss of generality.

\subsection{EAPs' Utilities}
The utility of the $k$th EAP is defined as
\begin{equation}\label{eq:UEAP}
U_k = \pi_k - \mathcal{C}_k (p_k),
\end{equation}
where $p_k = q_k /G_{k,s}$ is the transmit power of the $k$th EAP, and $\mathcal{C}_k(\cdot)$ is used to model the energy cost of the $k$th EAP, given by
\begin{equation}
\mathcal{C}_k (x) = a_k x^2,
\end{equation}
where $a_k > 0$. Note that the above quadratic function has been widely adopted in the energy trading market to model the energy cost\cite{Mohsenian2010Autonomous}. Equivalently, (\ref{eq:UEAP}) can be rewritten as
\begin{equation}\label{eq:UEAP2}
U_k = \pi_k - \frac{a_k}{G_{k,s}^2} q_k^2.
\end{equation}
We define the type of the $k$th EAP as
\begin{equation}\label{eq:theta}
\theta_k := \frac{G_{k,s}^2}{a_k},
\end{equation}
which suggests that the stronger the channel quality $G_{k,s}$ between the EAP and the information source, and/or the lower the unit power cost $a_k$, the higher the type of the EAP. Without loss of generality, we assume that there are totally $K$ types of EAPs with $\theta_1 < \theta_2 < \dots < \theta_K$. In this definition, the higher type EAP has better channel quality and/or lower energy cost. Note that since $a_k>0$ and $G_{k,s}>0$, $\theta >0$ holds. Using (\ref{eq:theta}), the EAP's utility can be rewritten as
\begin{equation}\label{eq:UEAPSim}
U_k = \pi_k-\frac{q_k^2}{\theta_k}.
\end{equation}
Assume there are $N_k$ EAPs belonging to the $k$th type, we thus have $\sum\nolimits_{k=1}^{K} N_k = N$. We then can rewrite the DAP's utility according to the types of EAPs as
\begin{equation}\label{eq:UDAPNew}
U_{DAP} = W \log_2 \left( 1 + \gamma \sum\limits_{k=1}^{K} N_k q_k \right) - \sum\limits_{k=1}^K N_k \pi_k.
\end{equation}

\subsection{Social Welfare}
The social welfare is defined as the summation of the utilities of the DAP and all $N$ EAPs, given by
\begin{equation}\label{eq:SocWel}
\begin{aligned}
  \Gamma &= U_{DAP} + \sum\limits_{k=1}^{K} N_k U_k \\
  &= W \log_2 \left( 1+ \gamma \sum\limits_{k=1}^K N_k q_k \right) - \sum\limits_{k=1}^K \frac{N_k q_k^2}{\theta_k}.
\end{aligned}
\end{equation}
It can be seen that the internal transfers, i.e., rewards, are cancelled in the social welfare, which is consistent with the aim to maximize the efficiency of the whole system, i.e., achieving more throughput at the cost of less energy consumptions.

\section{Contract Formulation}
In this section, we will formulate a contract for the energy trading between the DAP and EAPs, characterize its feasibility conditions, and derive the optimal contract subject to the feasibility conditions.

To overcome the information asymmetry, a contract including a series of energy-reward pairs $(q_k, \pi_k)$ (i.e., contract item) is designed to maximize the expectation of the DAP's utility, which is consistent with the social welfare in our model. For the $k$th type EAP, $q_k$ is the received power contributed by $k$th EAP\footnote{Note that the received power contributed by each EAP is assumed to be distinguishable by considering that the  EAPs work in disjoint narrow bandwidth. } and $\pi_k$ is the money paid to the $k$th EAP as the reward for its contribution.
\subsection{Optimal Contract with Asymmetric Information}
Generally, the first step in a contract design is to figure out its feasibility conditions. In our design, to encourage the EAPs to participate in the charging process and ensure that each EAP only chooses the contract item designed for its type, the following individual rationality (IR) and incentive compatibility (IC) constraints should be satisfied \cite{Bolton2005Contract}.

\emph{Definition 1: Individual Rationality (IR).} The contract item that an EAP chooses should ensure a nonnegative utility, i.e.,
\begin{equation}\label{eq:IR}
  U_k = \pi_k - \frac{q_k^2}{\theta_k} \ge 0, \forall k \in \{1,\dots,K\}.
\end{equation}

\emph{Definition 2: Incentive Compatibility (IC).} An EAP of any type $k$ prefers to choose the contract item $(q_k, \pi_k)$ designed for its type, instead of any other contract item $(q_j, \pi_j), \forall j \in \{1,\dots,K\}$ and $j \ne k$, given by
\begin{equation}\label{eq:IC}
  \pi_k - \frac{q_k^2}{\theta_k} \ge \pi_j - \frac{q_j^2}{\theta_k}, \forall k,j \in \{1,\dots,K\}.
\end{equation}

The IR condition requires that the received reward of each EAP should compensate the cost of its consumed energy when it participates in the energy trading. If $U_k \le 0$, the EAP will choose not to charge the information source for the DAP. We define this case as $(q_k = 0, \pi_k = 0)$. The IC condition ensures that each EAP automatically selects the contract item designed for its corresponding type. The type of each EAP is thus revealed to the DAP, which is called ``self-reveal''. If a contract satisfies the IR and IC constraints, we refer to the contract as a feasible contract.

In this paper, we consider a scenario with strong information asymmetry. In such a scenario, the DAP is only aware of the total number of EAPs (i.e., $N$) and the distribution of each type. But it does not know the exact number of each type $k$ (i.e., $N_k$). So the DAP needs to optimize its expected utility over the possibilities of all possible combinations of $N_k$. The expected utility of the DAP with $N$ EAPs is given by
\begin{equation}\label{eq:Expect_U_DAP}
\begin{aligned}
  &\mathbb{E}\{U_{DAP}\} = \sum\limits_{n_1=0}^{N}\sum\limits_{n_2=0}^{N-n_1}\dots \sum\limits_{n_{K-1}=0}^{N-\sum\nolimits_{i=0}^{K-2}n_i}\\
  &\left\{ \Phi_{n_1,\dots,n_K} \left[ W \log_2 \left( 1 + \gamma \sum\limits_{k=1}^{K} n_k q_k \right) - \sum\limits_{k=1}^K n_k \pi_k \right] \right\},
\end{aligned}
\end{equation}
where $\Phi_{n_1,\dots,n_K}$ is the probability of a certain combination of the number of EAPs belonging to each type (i.e., $N_k,\{k=1,2,\dots,K\}$) and $n_K = N - \sum\nolimits_{i=0}^{K-1}n_i$ is known after giving $n_1,n_2,\dots,n_{K-1}$ since the DAP knows the total number $N$ of EAPs. We assume that all types are uniformly distributed, so the probability of one EAP belonging to each type is the same, which is $1/K$. In this case, $\Phi_{n_1,\dots,n_K}$ can be calculated as
\begin{equation}\label{eq:Prob_combine}
\begin{aligned}
  \Phi_{n_1,\dots,n_K}&=\textbf{Pr} \left( N_1=n_1, N_2=n_2, \dots, N_{k}=n_{k} \right)\\
  &=\frac{N!}{n_1! n_2! \dots n_K! K^N}
\end{aligned}
\end{equation}
Recall that the DAP aims at maximizing its expected utility subjecting to the constraints of IR and IC given in (\ref{eq:IR}) and (\ref{eq:IC}). Thus, the optimal contract becomes
\begin{equation}\label{eq:CntFrm}
\begin{aligned}
  & \qquad \max\limits_{(q_k, \pi_k)}  \mathbb{E}\{U_{DAP}\} \\
  s.t. \quad & \pi_k - \frac{q_k^2}{\theta_k} \ge 0, \forall k \in \{1,\dots,K\}, \\
  &\pi_k - \frac{q_k^2}{\theta_k} \ge \pi_j - \frac{q_j^2}{\theta_k}, \forall k,j \in \{1,\dots,K\},\\
  &q_k \ge 0, \pi_k \ge 0, \theta_k \ge 0, \forall k \in \{1,\dots,K\}.
\end{aligned}
\end{equation}
The first two constraints correspond to IR and IC, respectively. Note that the EAP will reveal its private type truthfully with the IR and IC constraints. Specifically, the IR condition ensures the EAP's participation and the IC condition ensures that each EAP selects the contract item designed for its corresponding type to gain highest payoff.

\subsection{Problem Transformation}
There are $K$ IR constraints and $K(K-1)$ IC constraints in (\ref{eq:CntFrm}), which are non-convex and couple different EAPs together. It is hard to solve (\ref{eq:CntFrm}) directly due to the complicated constraints. Motivated by this, in the subsection we first reduce the constraints of (\ref{eq:CntFrm}) and transform it.

We first realize that the following necessary conditions can be derived from the IR and IC constraints.

\emph{Lemma 1:} For any feasible contract, $\pi_i > \pi_j$ if and only if $q_i > q_j$, $\forall i,j \in \{1,\dots,K\}$ and $\pi_i = \pi_j$ if and only if $\theta_i = \theta_j$, $\forall i,j \in \{1,\dots,K\}$.

\emph{Lemma 2:} For any feasible contract, $\pi_i = \pi_j$ if and only if $\theta_i = \theta_j$, $\forall i,j \in \{1,\dots,K\}$.

\emph{Lemma 3:} For any feasible contract, if $\theta_i > \theta_j$, then $\pi_i > \pi_j$, $\forall i,j \in \{1,\dots,K\}$.

Note that the proof for Lemma 1 to Lemma 3 are omitted due to the space limitation. Lemma 1 shows that the EAP contributing more received power at the information source will receive more reward. Lemma 2 indicates that the EAPs providing the same received power will get the same amount of reward. Lemma 3 shows that a higher type EAP should be given more reward. Together with Lemma 1 and Lemma 2, it can be duduced that a higher type EAP also contributes more energy to the information source. We define this feature as monotonicity.

\emph{Definition 3: Monotonicity.} If $\theta_i \ge \theta_j, \forall i,j \in \{1,\dots,K\}$ and then $\pi_i \ge \pi_j$.

Based on the above analysis, we can now use the IC condition to reduce the IR constraints and have the following lemma.

\emph{Lemma 4:} With the IC condition, the IR constraints can be reduced as
\begin{equation}\label{eq:ReIR}
  \pi_1 - \frac{q_1^2}{\theta_1} \ge 0.
\end{equation}
\begin{proof}
Due to the IC condition, we have
\begin{equation}\label{eq:ReIRP-1}
\pi_k - \frac{q_k^2}{\theta_k} \ge \pi_1 - \frac{q_1^2}{\theta_k}.
\end{equation}
Since we have defined that $\theta_1 < \theta_2 < \dots < \theta_K$, we have
\begin{equation}\label{eq:ReIRP-3}
\pi_k - \frac{q_k^2}{\theta_k} \ge \pi_1 - \frac{q_1^2}{\theta_1} \ge 0.
\end{equation}

Note that (\ref{eq:ReIRP-3}) shows that with the IC condition, if the IR condition of the EAP with type $\theta_1$ holds, the IR condition of the other $K-1$ types will also hold. So the other $K-1$ IR conditions can be bind into the IR condition of the EAP with type $\theta_1$.
\end{proof}

We can also reduce the IC constraints and attain the following lemma.

\emph{Lemma 5:} With monotonicity, the IC condition can be reduced as the
local downward incentive compatibility (LDIC), given by
\begin{equation}\label{eq:LDIC}
\pi_i - \frac{q_i^2}{\theta_i} \ge \pi_{i-1} - \frac{q_{i-1}^2}{\theta_{i}}, \forall i \in \{2,\dots,K\},
\end{equation}
and the local upward incentive compatibility (LUIC), given by
\begin{equation}\label{eq:LUIC}
\pi_i - \frac{q_i^2}{\theta_i} \ge \pi_{i+1} - \frac{q_{i+1}^2}{\theta_{i}}, \forall i \in \{1,\dots,K-1\},
\end{equation}

\begin{proof}
There are $K(K-1)$ IC constraints in (\ref{eq:CntFrm}), which can be divided into $K(K-1)/2$ downward incentive compatibility (DIC), given by
\begin{equation}\label{eq:DIC}
\pi_i - \frac{q_i^2}{\theta_i} \ge \pi_{j} - \frac{q_{j}^2}{\theta_{i}}, \forall i,j \in \{2,\dots,K\}, i>j,
\end{equation}
and $K(K-1)/2$ upward incentive compatibility (UIC), given by
\begin{equation}\label{eq:UIC}
\pi_i - \frac{q_i^2}{\theta_i} \ge \pi_{j} - \frac{q_{j}^2}{\theta_{i}}, \forall i,j \in \{2,\dots,K\}, i<j,
\end{equation}

Let's first prove the DIC can be reduced as the LDIC. By using the LDIC for three continuous types, $\theta_{i-1} < \theta_{i} < \theta_{i+1}, \forall i \in \{2,\dots,K-1\}$, we have
\begin{equation}\label{eq:LDICP-1}
\pi_{i+1} - \frac{q_{i+1}^2}{\theta_{i+1}} \ge \pi_{i} - \frac{q_{i}^2}{\theta_{i+1}},
\end{equation}
\begin{equation}\label{eq:LDICP-2}
\pi_i - \frac{q_i^2}{\theta_i} \ge \pi_{i-1} - \frac{q_{i-1}^2}{\theta_{i}}, \forall i.
\end{equation}
By applying the monotonicity, i.e., if $\theta_i > \theta_j$, then $\pi_i > \pi_j$, $\forall i,j \in \{1,\dots,K\}$, we have
\begin{equation}\label{eq:LDICP-4}
\theta_{i+1}(\pi_{i} - \pi_{i-1}) \ge \theta_{i}(\pi_{i} - \pi_{i-1}),
\end{equation}
Combine (\ref{eq:LDICP-2}) and (\ref{eq:LDICP-4}), we have
\begin{equation}\label{eq:LDICP-6}
\pi_{i} - \frac{q_i^2}{\theta_{i+1}} \ge \pi_{i-1} - \frac{q_{i-1}^2}{\theta_{i+1}}.
\end{equation}
Combine (\ref{eq:LDICP-6}) and (\ref{eq:LDICP-1}), we have
\begin{equation}\label{eq:LDICP-7}
\pi_{i+1} - \frac{q_{i+1}^2}{\theta_{i+1}} \ge \pi_{i-1} - \frac{q_{i-1}^2}{\theta_{i+1}}.
\end{equation}
So far, we have proved that type $\theta_{i+1}$ will prefer contract item $(q_{i+1},\pi_{i+1})$ rather than contract item $(q_{i-1},\pi_{i-1})$. By using (\ref{eq:LDICP-7}), it can be extended downward until type $\theta_{1}$, and thus all DIC holds.
\begin{equation}\label{eq:LDICP-8}
\pi_{i+1} - \frac{q_{i+1}^2}{\theta_{i+1}} \ge \pi_{i-1} - \frac{q_{i-1}^2}{\theta_{i+1}} \ge \dots\ \ge \pi_{1} - \frac{q_{1}^2}{\theta_{1}}, \forall i.
\end{equation}
So we conclude that with the monotonicity and the LDIC, the DIC holds. Similarly, we can prove that with the monotonicity and the LUIC, the UIC holds.
\end{proof}

The LDIC and the LUIC can be combined as shown in Lemma 6.

\emph{Lemma 6:} Since the optimization objective function is an increasing function of $q_k$ and a decreasing function of $\pi_k$, the above optimal problem can be further simplified as
\begin{equation}\label{eq:CntFrmRe2}
\begin{aligned}
  & \qquad \max\limits_{(q_m, \pi_m)}  \mathbb{E}\{U_{DAP}\}  \\
  s.t. \quad & \pi_1 - \frac{q_1^2}{\theta_1} = 0, \\
  &\pi_k - \frac{q_k^2}{\theta_k} = \pi_{k-1} - \frac{q_{k-1}^2}{\theta_k}, \forall k \in \{2,\dots,K\},\\
  &\pi_K \ge \pi_{K-1} \ge \dots \ge \pi_1, \\
  &q_k \ge 0, \pi_k \ge 0, \theta_k \ge 0, \forall k \in \{1,\dots,K\}.
\end{aligned}
\end{equation}

The proof of Lemma 6 is omitted due to the space limitation. We now solve the optimization problem (\ref{eq:CntFrmRe2}) to attain the optimal contract in the subsequent way: a standard method is first applied to resolve the relaxed problem without monotonicity and the solution is then verified to satisfy the condition of the monotonicity. By iterating the first and second constraints in (\ref{eq:CntFrmRe2}) and substituting $\pi_k, \forall k \in \{1,\dots,K\}$ into $\mathbb{E}\{U_{DAP}\}$, all $\pi_k, \forall k \in \{1,\dots,K\}$ are removed from the optimization problem (\ref{eq:CntFrmRe2}), which becomes
\begin{equation}\label{eq:finalprob}
\begin{aligned}
  &\max\limits_{q_k} \sum\limits_{n_1=0}^{N}\sum\limits_{n_2=0}^{N-n_1}\dots \sum\limits_{n_{K-1}=0}^{N-\sum\nolimits_{i=0}^{K-2}n_i} \Phi_{n_1,\dots,n_K}\\
  &\times \left[ W \log_2 \left( 1 + \gamma \sum\limits_{k=1}^{K} n_k q_k \right) \right.\\
  &- \left. \sum\limits_{k=1}^{K-1}
\left( \frac{1}{\theta_{k-1}} \sum\limits_{i=k}^{K}n_i
+ \frac{1}{\theta_{k}} \sum\limits_{i=k+1}^{K}n_i \right)q_k^2
-\frac{n_K}{\theta_K}q_K^2 \right],\\
s.t. &~~~~q_k \ge 0, \forall k \in \{1,\dots,K\}.
\end{aligned}
\end{equation}
Note that (\ref{eq:finalprob}) becomes a concave problem. So we can leverage standard convex optimization tools in \cite{Boyd2004Convex} to solve it to get $q_k$, and then $\pi_k$ can be calculated iteratively by the first two constraints in (\ref{eq:finalprob}). Moreover, monotonicity is met automatically when the type is uniformly distributed\cite{Bolton2005Contract}. So far, we have derived the optimal contract $(q_k,\pi_k)$, $\forall k \in \{1,\dots,K\}$, which can maximize the utility of the DAP and satisfy the constraints of IR and IC.

\section{Simulations and Discussions}
In this section, we first evaluate the feasibility of the proposed contract, and then demonstrate the performance of the proposed incentive mechanism. For the purpose of comparisons, another two incentive mechanisms are also simulated in this section. The performance of the optimal contract with complete information (i.e., the DAP knows exactly the types of the EAPs) is introduced as the upper bound. The other one is a linear incentive mechanism, in which the DAP sets a uniform price $P$ for unit quality of received energy that is optimized to maximize the utility of the DAP.

\begin{table}[!t]
\renewcommand{\arraystretch}{1.3}
\caption{System Settings}
\label{system_settings}
\centering
\scalebox{1}[0.9]{
\begin{tabular}{|c|c|}
\hline
Parameters                                      & Values        \\
\hline
Energy harvesting efficiency $\eta$             & 0.5           \\
\hline
Bandwidth $W$                                   & 1MHz          \\
\hline
Energy cost coefficient $a_m$                   & [0.1,1]        \\
\hline
$d_{m,s}$                                       & [5m,10m]        \\
\hline
$d_{a,s}$                                       & [15m,25m]       \\
\hline
Path-loss coefficient $\alpha$                  & 2             \\
\hline
Power attenuation at reference distance of 1m   & 30dB         \\
\hline
Noise power $N_0$                               & $10^{-8}mW$   \\
\hline
\end{tabular}
}
\end{table}

\begin{figure}[!t]
  \centering\scalebox{0.5}{\includegraphics{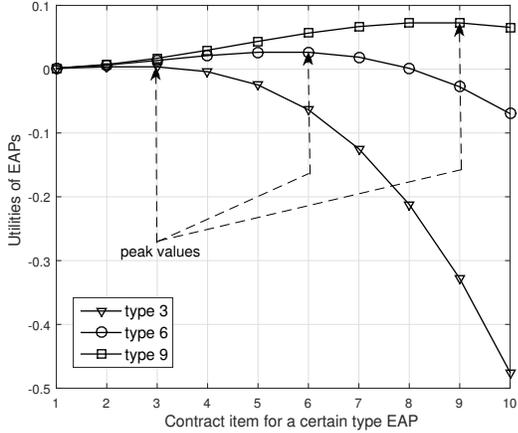}}
  \caption{Utilities of EAPs with type 3, type 6 and type 9 as functions of contract items designed for all kinds of EAPs from type 1 to type 10. We set $N=5$ and $K=10$.}\label{Fig:IC}
\end{figure}

The main system parameters are shown in Table I. Since $\theta=G_{m,s}^2/a_m$ and $\gamma=\eta G_{a,s}/N_0$, the practical ranges of $\theta$ and $\gamma$ can be determined by the parameters shown in Table I. The DAP's type $\theta$ is uniformly distributed. The unit of achievable throughput is set as Mbps.

To verify the feasibility (i.e., IR and IC) of the proposed scheme under information asymmetry, the utilities of EAPs with type 3, type 6 and type 9 are plotted in Fig. 2 as functions of all contract items $(q_k,\pi_k),k\in {1,2,\dots,K}$. It can be seen that each of the utility achieves its peak value only when it chooses the contract item designed for its corresponding type, which suggests the IC constraint is satisfied. For example, for the type 6 EAP, it achieves the peak value only when it selects the contract item $(q_6,\pi_6)$, which is designed for its type. If the type 6 EAP selects any other contract item $(q_k,\pi_k),k\in {1,2,\dots,K}$ and $k\ne 6$, its utility will be less than that when it selects the contract item $(q_6,\pi_6)$. Moreover, when each of above type EAPs (i.e., type 3, type 6 and type 9) chooses the contract item designed for its corresponding type, the utilities are nonnegative. Note that similar phenomenon can be observed for all other types of EAPs when they select the contract item designed for their corresponding types, which are not shown in Fig. 1 for brevity. So the IR condition is satisfied. It can be concluded that utilizing the proposed scheme, EAPs will automatically reveal its type to the DAP after its selection. This means that using the proposed scheme, the DAP can capture the EAPs' private information (i.e., its type), and thus overcome the problem of information asymmetry.

\begin{figure}[!t]
  \centering\scalebox{0.5}{\includegraphics{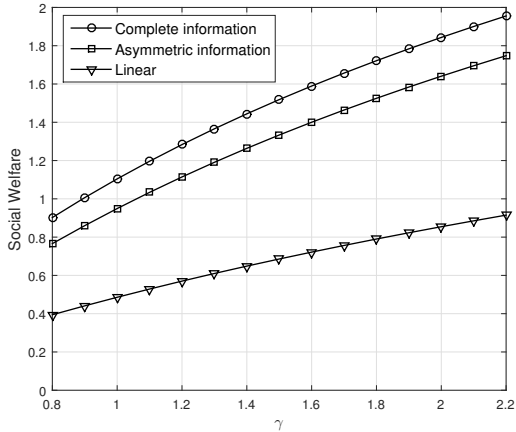}}
  \caption{Social welfare as a function of $\gamma$. We set $N=2$ and $K=5$.}\label{Fig:Social}
\end{figure}

\begin{figure}[!t]
  \centering\scalebox{0.5}{\includegraphics{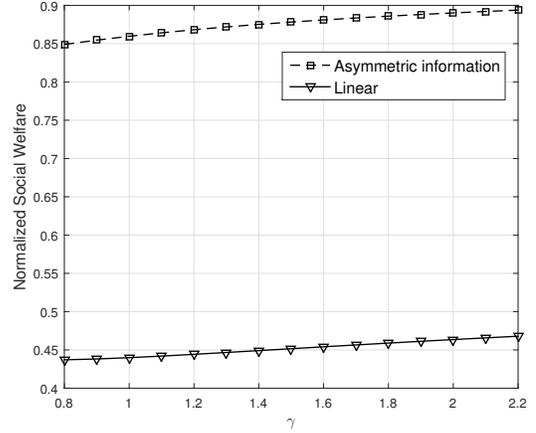}}
  \caption{Normalized social welfare as a function of $\gamma$. We set $N=2$ and $K=5$.}\label{Fig:SocialNormalized}
\end{figure}

%\begin{figure}[!ht]
%  % Requires \usepackage{graphicx}
%  \centering
%  \subfloat[social welfare]{\includegraphics[width=.24\textwidth]{fig2.eps}\label{Fig:SclWel}}
%  \subfloat[normalized social welfare]{\includegraphics[width=.24\textwidth]{fig3.eps}\label{Fig:SclNor}}
%  \caption{Social welfare and normalized social welfare as a function of $\gamma$. We set N=2 and K=5.}\label{Fig:SocialWelfare}
%\end{figure}

To evaluate the performance of the proposed scheme with asymmetric information, we compare its corresponding social welfare with those of the optimal scheme with complete information and another linear scheme with asymmetric information. Fig. 2 plots the curves of the social welfare as a function of $\gamma$. Fig. 3 shows the normalized social welfare as a function of $\gamma$, where social welfare of the proposed scheme and linear scheme are normalized by the social welfare of the optimal scheme with complete information. It can be observed from Fig. 2 that the utilities achieved by three of schemes all increase with $\gamma$. This is because with the same $\sum\nolimits_{m=1}^{N} q_m$, the larger the value of $\gamma$, the larger the achievable throughput $R_{sa}$ (refer to (\ref{eq:Ras2})), and thus larger social welfare (refer to (\ref{eq:SocWel})). Moreover, it is also shown in Fig. 2 that the performance of the optimal scheme with complete information providing the best performance serving as the upper bound, and the the performance of the linear scheme is the worst. It can be seen in Fig. 3 that the performance of the proposed scheme is generally larger that $85\%$ of that of the optimal scheme with complete information, and gradually approach to it with the increasing of $\gamma$. This demonstrates that the proposed incentive mechanism can effectively overcome information asymmetry by leveraging contract theory. While the performance of the linear scheme is generally less than $50\%$ of that of the optimal scheme with complete information. This is because the linear scheme does not utilize the private information (i.e., type) of the EAPs, and thus achieves a much lower social welfare.

\section{Conclusions}
We developed a contract theory based incentive mechanism for the energy trading in radio frequency energy harvesting (RFEH) based Internet of Things (IoT) systems. By providing compatible incentive to the energy access points (EAPs) under information asymmetry, the expected utility of the data access points (DAP) as well as social welfare is maximized. Moreover, the proposed mechanism can approach the performance of the optimal contract with complete information and significantly outperform the linear pricing-based approach.

%\section{Proof of the ...}

%\section{}

 \section*{Acknowledgment}
This work was supported in part by ARC grants FL160100032, DP150104019 and FT120100487. Zhanwei Hou's research is also supported by International Postgraduate Research Scholarship (IPRS) and Australian Postgraduate Award (APA).

\ifCLASSOPTIONcaptionsoff
  \newpage
\fi

\bibliographystyle{IEEEtran}
\bibliography{EH_Contract_Theory}

\begin{thebibliography}{10}
\providecommand{\url}[1]{#1}
\csname url@samestyle\endcsname
\providecommand{\newblock}{\relax}
\providecommand{\bibinfo}[2]{#2}
\providecommand{\BIBentrySTDinterwordspacing}{\spaceskip=0pt\relax}
\providecommand{\BIBentryALTinterwordstretchfactor}{4}
\providecommand{\BIBentryALTinterwordspacing}{\spaceskip=\fontdimen2\font plus
\BIBentryALTinterwordstretchfactor\fontdimen3\font minus
  \fontdimen4\font\relax}
\providecommand{\BIBforeignlanguage}[2]{{%
\expandafter\ifx\csname l@#1\endcsname\relax
\typeout{** WARNING: IEEEtran.bst: No hyphenation pattern has been}%
\typeout{** loaded for the language `#1'. Using the pattern for}%
\typeout{** the default language instead.}%
\else
\language=\csname l@#1\endcsname
\fi
#2}}
\providecommand{\BIBdecl}{\relax}
\BIBdecl

\bibitem{Gubbi2013IoT}
J.~Gubbi, R.~Buyya, S.~Marusic, and M.~Palaniswami, ``Internet of things
  ({I}o{T}): A vision, architectural elements, and future directions,''
  \emph{Future Generation Computer Systems}, vol.~29, no.~7, pp. 1645--1660,
  Sep. 2013.

\bibitem{Kama2015Wireless}
P.~Kamalinejad, C.~Mahapatra, Z.~Sheng, S.~Mirabbasi, V.~C. Leung, and Y.~L.
  Guan, ``Wireless energy harvesting for the {I}nternet of {T}hings,''
  \emph{IEEE Commun. Magazine}, vol.~53, no.~6, pp. 102--108, Jun. 2015.

\bibitem{Niyato2016Novel}
D.~Niyato, D.~I. Kim, P.~Wang, and L.~Song, ``A novel caching mechanism for
  {I}nternet of {T}hings ({I}o{T}) sensing service with energy harvesting,'' in
  \emph{Proc. ICC}, May. 2016, pp. 1--6.

\bibitem{Medepally2010Voluntary}
B.~Medepally and N.~B. Mehta, ``Voluntary energy harvesting relays and
  selection in cooperative wireless networks,'' \emph{IEEE Trans. Wireless
  Commun.}, vol.~9, no.~11, pp. 3543--3553, Sep. 2010.

\bibitem{Chen2015stackelberg}
H.~Chen, Y.~Li, Z.~Han, and B.~Vucetic, ``A stackelberg game-based energy
  trading scheme for power beacon-assisted wireless-powered communication,'' in
  \emph{Proc. ICASSP}, Apr. 2015, pp. 3177--3181.

\bibitem{ma2015distributed}
Y.~Ma, H.~Chen, Z.~Lin, Y.~Li, and B.~Vucetic, ``Distributed and optimal
  resource allocation for power beacon-assisted wireless-powered
  communications,'' \emph{IEEE Trans. Commun.}, vol.~63, no.~10, pp.
  3569--3583, Aug. 2015.

\bibitem{Bolton2005Contract}
P.~Bolton and M.~Dewatripont, \emph{Contract theory}.\hskip 1em plus 0.5em
  minus 0.4em\relax MIT press, 2005.

\bibitem{Zhang2015Contract}
Y.~Zhang, L.~Song, W.~Saad, Z.~Dawy, and Z.~Han, ``Contract-based incentive
  mechanisms for device-to-device communications in cellular networks,''
  \emph{IEEE J. Sel. Areas Commun.}, vol.~33, no.~10, pp. 2144--2155, May 2015.

\bibitem{Duan2014Cooperative}
L.~Duan, L.~Gao, and J.~Huang, ``Cooperative spectrum sharing: A contract-based
  approach,'' \emph{IEEE Trans. Mobile Commun.}, vol.~13, no.~1, pp. 174--187,
  Nov. 2014.

\bibitem{Krikidis2013harvest}
I.~Krikidis, G.~Zheng, and B.~Ottersten, ``Harvest-use cooperative networks
  with half/full-duplex relaying,'' in \emph{Proc. WCNC}, Apr. 2013, pp.
  4256--4260.

\bibitem{Ju2014Throughput}
H.~Ju and R.~Zhang, ``Throughput maximization in wireless powered communication
  networks,'' \emph{IEEE Trans. Wireless Commun.}, vol.~13, no.~1, pp.
  418--428, Dec. 2014.

\bibitem{Lu2015Wireless}
X.~Lu, P.~Wang, D.~Niyato, D.~I. Kim, and Z.~Han, ``Wireless networks with {RF}
  energy harvesting: A contemporary survey,'' \emph{IEEE Commun. Surveys
  Tuts.}, vol.~17, no.~2, pp. 757--789, Nov. 2015.

\bibitem{Mohsenian2010Autonomous}
A.-H. Mohsenian-Rad, V.~W. Wong, J.~Jatskevich, R.~Schober, and A.~Leon-Garcia,
  ``Autonomous demand-side management based on game-theoretic energy
  consumption scheduling for the future smart grid,'' \emph{IEEE Trans. Smart
  Grid}, vol.~1, no.~3, pp. 320--331, Nov. 2010.

\bibitem{Boyd2004Convex}
S.~Boyd and L.~Vandenberghe, \emph{Convex optimization}.\hskip 1em plus 0.5em
  minus 0.4em\relax Cambridge university press, 2004.

\end{thebibliography}

%\begin{IEEEbiography}{Yuguang ``Michael'' Fang}
%\end{IEEEbiography}

%It is not necessary to upload the biography when you submit your manuscript.

\end{document}